\documentclass[manuscript,nonacm]{acmart}
\usepackage{xcolor}

\AtBeginDocument{%
  \providecommand\BibTeX{{%
    \normalfont B\kern-0.5em{\scshape i\kern-0.25em b}\kern-0.8em\TeX}}}

\usepackage{xcolor, soul}

\soulregister{\cite}{7}
\soulregister{\ref}{7}

\begin{document}

\title[AI red-teaming is a sociotechnical problem: on values, labor, and harms]{AI red-teaming is a sociotechnical problem: on values, labor, and harms}

\author{Tarleton Gillespie}
\authornote{Tarleton Gillespie is a Senior Principal Researcher at Microsoft Research, Cambridge, Massachusetts, USA, and an affiliated Professor in the Department of Communication and the Department of Information Science at Cornell University, Ithaca, New York, USA}
\email{tarleton@microsoft.com}
\orcid{0000-0002-2601-6073}
\affiliation{%
 \institution{Microsoft Research}
 \city{Cambridge}
 \state{Massachusetts}
 \country{USA}}

\author{Ryland Shaw}
\authornote{Ryland Shaw is a doctoral student in the Annenberg School of Communication at the University of Southern California, and was a predoctoral research assistant at Microsoft Research, Cambridge, MA, USA at the time of this writing.}
\email{ryland.shaw@usc.edu}
\orcid{0009-0008-6123-1781}
\affiliation{%
 \institution{University of Southern California}
 \city{Los Angeles}
 \state{California}
 \country{USA}}

\author{Mary L. Gray}
\authornote{Mary L. Gray is a Senior Principal Researcher at Microsoft Research, Cambridge, Massachusetts, USA}
\email{mlg@microsoft.com}
\orcid{1234-5678-9012}
\affiliation{%
 \institution{Microsoft Research}
 \city{Cambridge}
 \state{Massachusetts}
 \country{USA}}

\author{Jina Suh}
\authornote{Jina Suh is a Principal Researcher at Microsoft Research, Redmond, Washington, USA}
\email{jinsuh@microsoft.com}
\orcid{0000-0002-7646-5563}
\affiliation{%
 \institution{Microsoft Research}
 \city{Redmond}
 \state{Washington}
 \country{USA}}



\begin{abstract}
As generative AI technologies find more and more real-world applications, the importance of testing their performance and safety is paramount. “Red-teaming” has quickly become the primary approach to testing AI models--prioritized by AI companies, and enshrined in AI policy and regulation. Members of red teams act as adversaries, probing AI systems to test their safety mechanisms and uncover vulnerabilities. Yet we know far too little about this work or its implications. In this essay we highlight the importance of understanding the values and assumptions behind red-teaming, the labor arrangements involved, and the psychological impacts on red-teamers, drawing insights from lessons learned around the work of content moderation. Red-teaming should be a deeply interdisciplinary concern. To avoid repeating the mistakes of the recent past, we call for a coordinated network of scholars, from the full range of the computational and social sciences, to study the technical, social, critical, and policy dimensions of red-teaming and of the emerging sociotechnical system that is AI.  
\end{abstract}



\keywords{AI safety, red team, sociotechnical systems, content moderation, labor, well-being}



\maketitle

Large language models (LLMs) and image diffusion models have emerged so rapidly, from research projects into the engines behind the global deployment of generative AI, whether accessed directly on the web or embedded in software, search, or social media. When a technology jumps this quickly from experimental plaything to widely available consumer service, many other elements also settle in around it, often without much forethought: interfaces, policies, business models, labor arrangements, infrastructural assurances, complementary technologies, public claims, advertising campaigns, regulations. Many of these decisions, arrangements, and infrastructures may turn out to be just as consequential for users and the broader public as the core technology itself. But the boisterous promises and debates that surround the new technology can obscure these other essential elements, elements that make technologies always more than the sum of their engineered parts. 

Researchers studying the workings and implications of these technologies, across computer science, engineering, the social sciences, humanities, and law, must gear up just as quickly, to study not just the core technology, but the sociotechnical system taking shape around it~\cite{joyce_toward_2021}.  And they need to study it together. In this essay, we call upon computer scientists and social scientists alike to pay closer, critical, and collaborative attention to one part of AI development, ``red-teaming.''\footnote{The spelling of the term is inconsistent across different texts and organizations (sometimes even within organizations): sometimes it is hyphenated or concatenated, other times not. In line with~\cite{friedler_ai_2023}, we hyphenate the term except when specifically referencing the teams themselves.} 

AI models and their applications typically undergo internal testing before release, and continue to be evaluated during use; one part of this testing, red-teaming, probes these applications for exploitable vulnerabilities, errors, and bias. From an AI evaluation vantage point, red-teaming needs to be well-designed, effective, and replicable. But from a sociological vantage point, red-teaming is something else as well: a specific kind of labor, done by specific sets of people, in specific institutional contexts, with its own specific set of implications~\cite{friedler_ai_2023, metcalf_scaling_2023}. Our aim is to shed light on these sociotechnical elements, and to call for more cross-disciplinary attention to this critical component of AI development. 

Since the commercial launch of ChatGPT, red-teaming has been quickly normalized as a step in the production and deployment of generative AI models. AI developers champion it as proof of their public responsibility, while regulators count on it as a bulwark preventing AI from inflicting social harms. But the public knows precious little about how this work is conducted, upon what values and assumptions it is based, who is enlisted to do it, or the psychological costs they bear. This was also the case with content moderation: too long hidden from public and critical scrutiny, its labor and well-being concerns were too long overlooked; and the opacity of these value judgments soon became a political liability for social media platforms. And, given that what little the public does know about AI red-teaming comes largely from Silicon Valley's own promotional materials, and given that the changing political climate particularly in the U.S. may discourage such outward performances of responsibility, the public may soon know even less.

It is worth noting that this essay is not based on any information internal to Microsoft. Rather, we share our observations, drawn from our recent studies of the labor behind Responsible AI \cite{gray2019ghost, zhang_human_2024, zhang2024aura}, the politics of generative AI as a new media technology \cite{gillespie_generative_2024},  and participatory approaches to AI development and governance \cite{suresh_participation_2024, deng_responsible_2024}.\footnote{Also see Microsoft Research's Project Resolve: https://www.microsoft.com/en-us/research/project/project-resolve/} And, being trained in science and technology studies, labor, psychology and design, we feel it is important to contextualize AI red-teaming within longer histories of scholarship that reckon with the social construction of new technologies.
 
\section{What is red-teaming?}

Broadly, ``red-teaming'' means testing the safety and security of a system by methodically probing it as an adversary would. The U.S. military coined the term to describe the technique of assigning members of one's own forces to act as the enemy during wargames and simulations, probing defensive strategies for potential weaknesses. During the Cold War, that presumed enemy was the Soviet Union, hence the color ``red''~\cite{zenko_red_2015}. But Zenko suggests that the idea of adversarial testing extends back much farther, noting that in the thirteenth century the Catholic church established the ``Devil's Advocate,'' who interrogated those nominated for sainthood. Just as the Devil's Advocate aimed to poke holes in nominees' candidacies, military red teams attempted to infiltrate their own forces' front lines. 

The term migrated to the field of cybersecurity. Red teams were tasked with infiltrating information systems to simulate worst case scenarios, like the theft of sensitive information or hacks on infrastructure, that might lead to financial or operational disaster~\cite{zenko_red_2015}. Red teams became an important element of systems security, that might even pay for themselves by anticipating and thereby preventing breaches.\footnote{see \url{https://www.wired.com/story/microsoft-ai-red-team/}} 

For generative AI, red-teaming sometimes concerns the security of the foundation model, but increasingly it also means purposefully provoking the model to produce undesired or incorrect responses. Because users might either intentionally or inadvertently prompt an AI model to generate hateful, pornographic, vile, or biased responses, red teams attempt to do so first, acting as a kind of adversary to the intended or presumed use. By preemptively tempting the AI model to say things it shouldn't, they can document how to shore up the safety architectures meant to prevent such responses. 

Still, what constitutes AI red-teaming remains fuzzy, given that its tactics and organizational structures are still forming. Its familial resemblance to evaluation, social engineering, bug bounties, threat assessments, and penetration testing is still "being discovered"  \cite{singh_red-teaming_2025}. Other terms from cybersecurity and hacker lingo describe much the same project: Google describes their AI Red Team as ``ethical hackers'', a nod to venerable ``white hat'' hackers-for-hire who hunt for technical insecurities.\footnote{\url{https://blog.google/technology/safety-security/googles-ai-red-team-the-ethical-hackers-making-ai-safer/}}

Some assert that red-teaming is essential to AI, and the surest way to safeguard equitable and responsible AI development. Others worry that red-teaming and the guardrails it produces are a kind of ``security theater,'' more performative than substantive~\cite{feffer2024red}, meant to obscure the reckless deployment of harmful technologies to the public. Others suggest that such efforts to police AI models will hinder the true potential of the technology, ultimately leading to ``woke AI''.\footnote{On X, Elon Musk claimed a ``woke AI'' may eventually kill people: \url{https://x.com/elonmusk/status/1768746706043035827}} 

But in blunt economic terms, the first AI company to successfully ``tame'' their generative AI products through such safeguards could well capture the market. Business clients are asking for AI-powered customer service chatbots that do not hallucinate\footnote{In 2024, Air Canada was found liable for its customer support chatbot's incorrect claim about a customer's ticket: \url{https://www.cbc.ca/news/canada/british-columbia/air-canada-chatbot-lawsuit-1.7116416}} and productivity tools that behave consistently.\footnote{Complaints about AI tools getting ``lazy'' surface from time to time: \url{https://www.theguardian.com/commentisfree/2024/jan/12/chatgpt-problems-lazy}, as well as concerns about generative AI systems degrading over time: \url{https://www.nytimes.com/interactive/2024/08/26/upshot/ai-synthetic-data.html}} There are enormous financial, reputational, and regulatory incentives to make generative AI tools safe and value-transparent, pushing these companies to rapidly institute red teams—arguably, faster than researchers concerned about the politics of AI can follow.  

Red-teaming also offers a kind of reassurance, easing fears held by the public, governments, and financial stakeholders about the safety and performance of generative AI systems. New AI products are often touted as having been tested by red-teamers before a wider release.\footnote{For example, when OpenAI announced its Sora video generation model in early 2024, they prominently featured their red-teaming efforts on its launch page. \protect{\url{https://web.archive.org/web/20240215192216/https://openai.com/sora\#safety}}} The US federal government (briefly) adopted the language of red-teaming as an important assurance of the safety of AI systems. In October 2023, US President Biden issued an Executive Order that reinforced the importance of red-teaming as part of a proposed system of federal oversight over AI: ``Any foundation model that poses a serious risk to national security, national economic security, or national public health and safety... must share the results of all red-team safety tests.''\footnote{White House Executive Order (14110) on the Safe, Secure and Trustworthy Development and Use of Artificial Intelligence, 2023} That order has since been rescinded, but red-teaming will likely appear in subsequent efforts to regulate, or self-regulate, generative AI.

As a new labor formation, developing under financial and political pressure, red-teaming echoes other contingent forms of digital labor: data labeling, content moderation, enrichment services of all kinds--all arrangements for supporting the semi-automation of human judgment critical to data-driven technical systems. If we want AI that is not only safe and secure but also sustainable, we need to also study the labor arrangements emerging that are critical to it: the mix of internal teams testing out AI products, volunteers convened at hack-a-thon-like events, third party crowdwork vendors, and professional security firms. And, as we learned in the case of social media, we must attend to the psychological costs of the work asked of red teams charged with making AI safe, secure, and useful to everyone. 

Rather than focus exclusively on implementing and improving it, for the sake of the model, we need to better understand red-teaming as a practice, and understand its place in the development of generative AI tools, for the sake of the people involved. A sociological perspective, which is fundamentally human-centric, can better guide the responsible and effective use of red teams.

\section{Value Judgments}

Efforts to insert obligations, structures, and benchmarks have had to race to keep up with the rush to commercialize generative AI models \cite{lazar_ai_2023}. Initial red-teaming efforts were being implemented even as design teams were still wondering which harms to even probe for: AI red-teamers have had to develop homegrown taxonomies of harms, and the measurement and benchmarking systems for mitigating them~\cite{hao_safety_2023, weidinger_sociotechnical_2023}. In public statements, the major AI companies often state as a given that generative AI tools will unavoidably produce harmful content~\cite{marres_ai_2024}. However, it is much rarer for them to discuss how they determine what counts as harmful content, what they should and should not be looking for, and whether their own teams are best suited to make those judgments. This prompts the question, ``whose values are being utilized for alignment and evaluation?'' \cite{feffer2024red}

The development of AI red-teaming echoes the early days of commercial content moderation at social media platforms.\footnote{\url{https://www.techpolicy.press/ais-content-moderation-moment-is-here/}}  The parallels are revealing. The categories of concern are strikingly similar:  graphic violence, hate speech, harassment, discrimination, sexual content, terrorism, human trafficking, self-harm, child abuse, and misinformation ~\cite{ganguli_red_2022,OpenAI_2023-GPT4_Systemcard}. And when Silicon Valley found itself compelled to manage the gap between what can be generated online and what users should actually see, it too enlisted contingent human labor to serve as that filter. 

Social media platforms ``discovered'' the need for human moderation labor after being surprised by the kinds of content that could turn up through their services~\cite{gillespie_custodians_2018, klonick_new_2018,gray2019ghost}. This awareness often came from user complaints, the technology press calling out the platforms' shortcomings, and platform companies stumbling upon it themselves. AI red-teaming has similarly been fueled by user complaints and critical press coverage. Solving these harms has to be done internally and largely in proprietary ways~\cite{carmi_hidden_2019, roberts_commercial_2016}, making industry-wide or public-wide discussions about harms and values difficult to develop. But leaving the public out of this process leaves the value judgments to the AI designers themselves. Or as OpenAI explained, echoing so many social media companies before them, ``Our approach is to red-team iteratively, starting with an initial hypothesis of which areas may be the highest risk, testing these areas, and adjusting as we go.''~\cite{OpenAI_2023-GPT4_Systemcard} 

Tackling harmful content internally, intuitively, and iteratively had profound implications for social media platforms over the past two decades; the same implications could befall the red-teaming of generative AI systems. Like AI red teams today, many social media platforms turned first to their own engineers and employees to evaluate for harmful content; to them, some categories of harms tended to seem more obvious, others tended to go unrecognized. Silicon Valley engineers generally do not reflect the range of identities and contexts of their global user base; social media platforms that began with their own employees often underestimated the harms faced by women, marginalized racial and ethnic groups, and those with stigmatized sexual and gender identities ~\cite{gerrard_content_2020, roberts_commercial_2016, ruckenstein_re-humanizing_2019, shahid_decolonizing_2023}.

In the earliest days of social media platforms, the predominant approach to moderation was the ``Feel bad? Take it down'' rule~\cite{klonick_new_2018}. Subjective judgments and gut feelings stood in, often poorly, for the public's understandings of harms or ethical principles. AI red-teaming strategies will have to be more complex and more inclusive to avoid the mistakes of social media's past. However, with little opportunity for outside researchers to study commercial red teams, and little internal or external incentive to disclose much about their practices, we lack a clear empirical understanding of what harms may be under-attended to, or fall outside the purview of a commercial organization.

\section{Labor politics}

The failure to appreciate the importance of human labor in AI systems, whether intentional or not, is common~\cite{gray2019ghost,roberts_behind_2019}. To this point, it is illustrative that `red-teaming' is often referred to as a verb, eliding the human workforce that constitutes red teams. A sociotechnical examination of red-teaming should extend not only to the values and concerns behind the techniques red teams deploy, but to the people doing the work and the labor arrangements within which they operate. 

Red-teaming as a method is emerging in various forms: inside and outside companies, salaried and volunteer, with access to the inner workings of the AI system and without.  These are bound to change: some forms will fall away, while others will settle in as ``the way things are done.'' But whatever particular labor politics do settle into place, there are important questions about the institutional contexts, material conditions, and economic incentives of this AI-related work, ripe for scholarly analysis.

Who does this work internally can vary. Big tech companies are eager to boast about their flagship red teams,\footnote{Google even produced a short documentary on their premier red-teamers that has been viewed over 1.5 million times as of December 2025: https://youtu.be/TusQWn2TQxQ?si=UnHO87JYDk1dblvK} whose jobs are dedicated solely to ethical hacking, but we know that red-teaming also happens at smaller scales throughout the product development cycle \cite{singh_red-teaming_2025}.  People doing the red-teaming may be part of larger Responsible AI efforts, Trust \& Safety divisions, or legal/compliance apparatuses \cite{zhang2024aura}. While those who perform red-teaming for their own companies typically enjoy the job security of full employment, they may not be in a position to refuse a red-teaming request. While they are likely to have the necessary technical understanding of how models work, it is not clear that this is sufficient to effectively identify and mitigate AI risks \cite{singh_red-teaming_2025}. And they may not be able to raise concerns publicly without breaching corporate norms or legally binding non-disclosure agreements~\cite{christin_internal_2024}. 

Employee red-teamers may have little training in any other relevant proficiencies, whether linguistic, sociocultural, historical, legal, or ethical; the incentive structures do not ask for or reward such expertise. Some internal red teams might include someone with sociocultural domain expertise, but they, too, work for the company and may have conflicting incentives~\cite{friedler_ai_2023}. Those windows of opportunity can open, a little. To test GPT-4 ahead of its March 2023 release, OpenAI solicited help from more than 50 experts, though primarily from Trust and Safety and cybersecurity backgrounds~\cite{OpenAI_2023-GPT4_Systemcard}. Anthropic and Microsoft encourage their red teams to consult with experts to test specific types of harms~\cite{Microsoft_2023}. These partnerships give AI companies access to experts without having to retain them as formal employees, compensating them in clout, API credits, job opportunities, or bragging rights rather than dollars. 

Following a pattern ubiquitous in Silicon Valley, there have been increasing efforts to shift red-teaming labor from company employees to third-party datawork vendors, often overseas. Doing so uses labor arbitrage to drive down wages and, as the history of outsourcing and offshoring demostrates, can erode worker protections and their capacity to raise concerns about the effectiveness of their work. Early in their model development, researchers at Anthropic enlisted several hundred untrained crowdworkers and instructed them to ``make the AI behave badly'' to elicit harmful responses from their LLM chatbot~\cite{ganguli_red_2022}, In this project, the crowdworkers were responsible for both probing the AI and assessing its responses for harmful content. 

Either way, red teams can be an expensive undertaking. Smaller companies competing to bring their generative AI services to market may not be equipped to employ internal red-teaming services sufficient to satisfy internal liability concerns or regulatory obligations. They may turn to an emerging crop of boutique AI safety and data services\footnote{Such as Noma, Hidden Layer, Protect AI and Mindgard}  that see market opportunities in sourcing red-teaming. As scholars have noted~\cite{gray2019ghost}, when labor is offloaded, particularly piecemeal to a globally-distributed contingent workforce, it becomes harder to trace and trickier to assert labor protections for it.

Some red-teaming happens outside the confines of AI companies and their outsourced labor pools. Hackers, volunteers, and everyday users also engage in forms of red-teaming. At DEFCON 2023, for instance, one of the largest hacking conferences in the world, over 2000 volunteers came together to prompt the largest LLMs into producing harmful content.\footnote{\url{https://cyberscoop.com/def-con-ai-hacking-red-team/}} Attendees included field experts, some industry-based red-teamers, cybersecurity consultants, and CS PhD students, but DEFCON organizers also dismantled some significant technical barriers to broaden participation to include novices, even children, with no programming knowledge~\cite{cattell_generative_2023}. 

The event demonstrated the viability of convening a greater diversity of perspectives than is represented on most internal company red teams, something AI ethicist Rumman Chowdhury has argued is imperative when red-teaming on issues such as race, gender, sexuality, politics, and class~\cite{oremus_ai_2023}. However, asking marginalized communities to elaborate on their own marginalization for red-teaming can easily slip into a transactional, extractive, and exploitative codependency \cite{dalal_provocation_2024}. Volunteer and uncompensated work also requires significant effort to organize and has no obvious scaling mechanism. Access to generative AI models must be negotiated with AI companies, and taking any action on the insights gleaned depends entirely on the companies' goodwill.

In many of the major generative AI applications, end users can also provide feedback, at least in limited forms. ChatGPT's ``thumbs down'' includes a pull-down menu for users to indicate their concern: ``Don't like the style,'' ``Not factually correct,'' ``Refused when it shouldn't have,'' and others. But it is not evident how feedback from these slivers of digital ``civic labor''~\cite{matias_civic_2019} ever gets back to designers. User feedback is not strictly speaking red-teaming, as it isn't necessarily adversarial. But end users do, whether inadvertently, playfully, or deviously, help designers surface outputs that companies did not anticipate or design for.

Red-teaming is still taking shape as a set of labor practices, inside and outside of AI companies. It is unclear whether or not companies' internal red-teaming efforts are limited by the diversity of their employee pool, and how often they lean unfairly on their own minoritized employees to inject some diversity into their models~\cite{benjamin_race_2019, lazar_ai_2023}.  And, outsourced red-teaming work and using labor arbitrage to reduce costs raises tragically familiar concerns that echo social media companies' use of human labor: leaving workers with fewer labor protections, more adverse work conditions, and more precarious job security, with few or no avenues for career development.

\section{Well-being of Red-Teamers}

Beyond understanding \textit{who} is doing the AI red-teaming and \textit{what} is being evaluated, we also need to pay attention to the human \textit{cost} of doing such work. Scholars and practitioners involved in red-teaming call it ``rather unsavory work''~\cite{bai_constitutional_2022} and ``mentally taxing''~\cite{Microsoft_2023}. Like content moderation work, the adversarial probing of red-teaming requires workers to imagine worst-case scenarios and expose themselves to potentially troubling outputs. Red-teamers are often required to assume the persona of a potential adversary (for example: online harasser, sex trafficker, racist, terrorist), invent a plan that this adversary may use to compromise the system, and evaluate the output for potential harms. They may also assume the persona of a benign user with specific intents or contexts (for example, a user with a history of eating disorders looking for dieting advice) and try to reveal system vulnerabilities that might be harmful. A red-teamer may first research harmful groups to learn their behaviors and bring that knowledge into red-teaming the models for hate speech or deep fakes. They may immerse themselves in child online safety concerns to then evaluate the model's capabilities in aiding child exploitation. 

To date, there is little empirical research about the psychological impact of AI red-teaming. But one recent study of RAI content workers shows that red-teamers, content moderators, and data labelers face similar psychological challenges when working with potentially harmful content~\cite{zhang2024aura}.
Given the extensive research on content moderation and the well-documented occupational health concerns experienced by professions that contend with trauma exposure~\cite{Steiger2021ThePW}, red-teamers stand to benefit from attention to this history.

Moreover, the success of a red-team operation depends on uncovering and reviewing increasingly harmful content. Much like content moderators, emergency responders, journalists, or police investigating distressing events, AI red-teamers may be exposed repeatedly to disturbing and traumatic content that can lead to negative psychological symptoms. For example, content moderators removing harmful and offensive material from platforms have reported mental health challenges extending long after the work is complete, leading to documented cases of post-traumatic stress disorder (PTSD) and secondary traumatic stress (STS)~\cite{ruckenstein_re-humanizing_2019, arsht_2018_human}. Prolonged exposure may contribute to long-lasting mental health symptoms, alterations to their personal belief systems, and increased risks of physical health issues and substance abuse~\cite{Steiger2021ThePW}. In fact, repeated work-related exposure to traumatic content is among the diagnostic criteria used for PTSD in the DSM5 (Diagnostic and Statistical Manual of Mental Disorders) and has subsequently been used to support a series of recent lawsuits brought by content moderators against their employers~\cite{pinchevski_social_2023}.\footnote{See \url{https://www.nytimes.com/2018/09/25/technology/facebook-moderator-job-ptsd-lawsuit.html}} Although many major platforms were slow to deal with this problem, most now offer mental health support and take measures to limit moderators' exposure to the most reprehensible content. These parallels underscore the importance of initiating research to protect red-teamers from the psychological hazards inherent in their work.

At the same time, AI red-teaming introduces distinct psychological challenges. A successful AI red-teamer must exhibit an antagonistic imagination to be effective. Or as one red-teamer put it: ``If there were a red-team motto, it would be: The more sinister your imagination, the better your work.''\footnote{\url{https://www.bostonglobe.com/2024/01/11/opinion/ai-testing-red-team-human-toll/}} Red-teaming involves deliberately engaging in transgressive, uncomfortable, unethical, immoral, or harmful activities, including immersing themselves in scenarios that go against their morals or belief systems--to think like a harasser, or feel like a target of discrimination. Such practice can lead to ``moral injury''~\cite{shay2014moral}, a form of psychological distress that stems from actions, or the lack thereof, that violate one's moral or ethical code~\cite{zhang2024aura}. Those who cannot safely detach their personal identity from their transgressions may experience negative self-perception and guilt. Regularly breaking the rules for the greater good can lead to a ``loss of self,'' sometimes seen in the undercover police profession~\cite{Joh2009BreakingTL}. 

The potential negative impacts on red-teamers' wellbeing have been acknowledged by some of those who organize such work. For example, organizers of the DEFCON Generative Red Team event anticipated that models generating unexpected harmful outputs might be triggering to participants~\cite{cattell_generative_2023}. Anthropic's early red-teaming efforts involved consultations with Trust \& Safety professionals to design safety measures for their crowdworkers~\cite{ganguli_red_2022}. Strategies for preserving the wellbeing of red-teamers could include providing warnings about sensitive content, allowing opt-outs, encouraging breaks, monitoring mood, or allowing them to choose topics within their own risk tolerance~\cite{ganguli_red_2022, OpenAI_2023-GPT4_Systemcard, Microsoft_2023}. But for volunteer or crowdsourced red-teamers, such strategies are limited to what the organizers are willing to provide.

For professional red-teamers, companies sometimes offer employee assistance programs (EAPs) with mental health resources. However, organizational factors often impact how these resources are actually used. For some, accessing a therapist may be a luxury they cannot afford, because of a psychologically unsafe work environment, unrelenting performance metrics, or job insecurity. Non-disclosure agreements (NDAs) have historically prevented workers from speaking up about working conditions~\cite{Steiger2021ThePW}. Monitoring red-teamers' wellbeing gets entangled with more unsavory forms of workplace surveillance~\cite{ajunwa2017limitless}, and increases liability risks for employers if the findings are severely negative. And implementing consistent wellbeing strategies can become infeasible for red-teamers working across organizational and national boundaries. So when organizers claim red-teaming events were safe because no one used the on-call therapists~\cite{cattell_generative_2023}, they may be mistakenly assuming that no usage means no need. 

Beyond individual mental health resources, one of the most effective tools for red-teamer wellbeing may be social support through a community of workers in similar roles, as well as family and friends. Prior research on content moderators demonstrates that the validation and belonging that comes from social support are essential~\cite{Steiger2021ThePW}. Informal red team communities~\cite{inie2023summon} that have emerged on social networks to share strategies can act as safe spaces to heal from shared trauma, especially when red-teamers may be reluctant to share their experiences with loved ones to protect them from exposure~\cite{Steiger2021ThePW}. 

In an attempt to minimize human exposure to the ``unsavory work'' of red-teaming, some researchers have advocated for automated red-teaming. Whether or not this is possible--some leading AI safety researchers have recently argued that the "human element" will always be necessary \cite{bullwinkel_lessons_2025}--automation may inadvertently make the human work even more invisible~\cite{gray2019ghost}, diverting resources away from well-being measures needed by those red-teamers who remain. Therefore, it is important to critically examine the role of automation in red-teaming—not only its immediate impact on tasks, but also its long-term effects on the overall ecosystem. 

We must acknowledge the sobering reality that comes with the commoditization of AI harm reduction. As long as generative AI remains integral to our lives, the work of AI red-teaming and its psychological implications will persist. To support this workforce, it is crucial to rigorously study and validate the effectiveness of innovative well-being strategies across various contexts, with close examination of the surrounding organizational and social structures.

\section{Conclusion}

What would more substantive, empirical, and cross-disciplinary research on red-teaming provide? Today's siloed, firewalled, market-reactive approach to red-teaming has potential drawbacks for AI consumers, red-teamers, and companies. Each company is rapidly developing its own version of red-teaming, with definitions and workloads varying based on the company's priorities, `brand,' and particular focus. Almost every company working on generative AI today has a red team workforce of some kind. While there is energy being put toward addressing biases and other ``embedded harms,'' plenty of red-teaming efforts are more concerned with ungrounded, existential risks than with current, tangible concerns. And it is unclear how many issues identified through red-teaming efforts have been mitigated. 

We must study how red-teaming judgments are made and by whom if we hope to improve its outcomes as part of a sociotechnical system. Its compartmentalization and operational opacity can both alienate workers and keep the public from understanding fully what AI systems can offer. Empirical study can challenge these arrangements, identify the barriers and incentives at play, and perhaps point the public and AI companies toward more sustainable alternatives.

An organized, crossdisciplinary, empirical research agenda that examines red-teaming as a sociotechnical undertaking could make both AI companies and the public more aware of and intentional about red-teaming work. Regardless of its current effectiveness or its future improvements, red-teaming has underlying logics and structural conditions that need examination. Lessons from content moderation show that finding and removing "bad elements" demands the deliberations and value judgments of teams of people. The specific conditions under which people do this hard work, at an unprecedented global scale and across myriad institutional settings, matters for both red team workers' occupational health and for the integrity of our technical and informational ecosystems.  

And like most data work, the notion that this labor is only temporary, that it will soon be automated away, is wrong, and (deliberately) distracting from these sociological concerns. It is difficult to believe that we can ever fully automate such peculiarly human judgments, about contentious and shifting topics, under pressure from regulators and the public, whose ethical frameworks can themselves shift. But even if it could be automated in the future, real people are doing this work right now, and with real consequences. It makes little difference whether we discard this phantasmic notion of full automation entirely, or just concede that data labor will be with us for the foreseeable future. Either way, research today can help structure this work in ways that are better attentive to the well-being and labor rights of the people doing the work right now.

In fact, we may need not just more empirical study of red-teaming, but a coordinated network of scholars studying red-teaming: as a multi-faceted practice, as a component in the institutional and labor arrangements of Silicon Valley, as a global public health concern, and as a hidden value system buried in our newest tools of expression and knowledge. The field of Computer Science has, in the last decade, begun to recognize that information systems are also labor systems and value systems--growing networks like the ACM Conference on Fairness, Accountability, and Transparency illustrate this--and it is grappling with the implications of that in ways it had not before. Again, we might learn from the rise of content moderation and the research that attended to it: many excellent scholars studying content moderation challenged its underlying logics and structural conditions. But, perhaps, a coordinated network of scholars to deepen, circulate, and affirm those insights could have had a more substantive impact on these arrangements. It is not too late to pose an empirical and coordinated challenge to red-teaming, and to the many forms of labor and values on which AI systems depend.


\end{document}